# A Manifesto for Web Science @10

by Wendy Hall[1], Jim Hendler[2], Steffen Staab[1,3]

[1]University of Southampton, wh@ecs.soton.ac.uk / s.r.staab@soton.ac.uk
[2]Rensellaer Polytechnic Institute, hendler@cs.rpi.edu
[3]Universität Koblenz-Landau, staab@uni-koblenz.de

Twenty-seven years ago, one of the biggest societal changes in human history began slowly when the technical foundations for the World Wide Web were defined by Tim Berners-Lee. Ever since, the Web has grown exponentially, reaching far beyond its original technical foundations and deeply affecting the world today – and even more so the society of the future.

We have seen that the Web can influence the realization of human rights [Wagner, 2016] and even the pursuit of happiness[1]. The Web provides an infrastructure to help us to learn, to work, to communicate with loved ones, and to provide entertainment. However, it also creates an environment affected by the digital divide between those who have and those who do not have access. Additionally, the Web provides challenges we must understand if we are to find a viable balance between data ownership and privacy protection, between over-whelming surveillance and the prevention of terrorism. For the Web to succeed, we need to understand its societal challenges including increased crime, the impact of social platforms and socio-economic discrimination, and we must work towards fairness, social inclusion, and open governance.

Ten years ago, the field of *Web Science* was created to explore the science underlying the Web from a socio-technical perspective including its mathematical properties, engineering principles, and social impacts [Berners-Lee et al, 2006]. Ten years later, we are learning much as the interdisciplinary endeavor to understand the Web's global information space continues to grow.

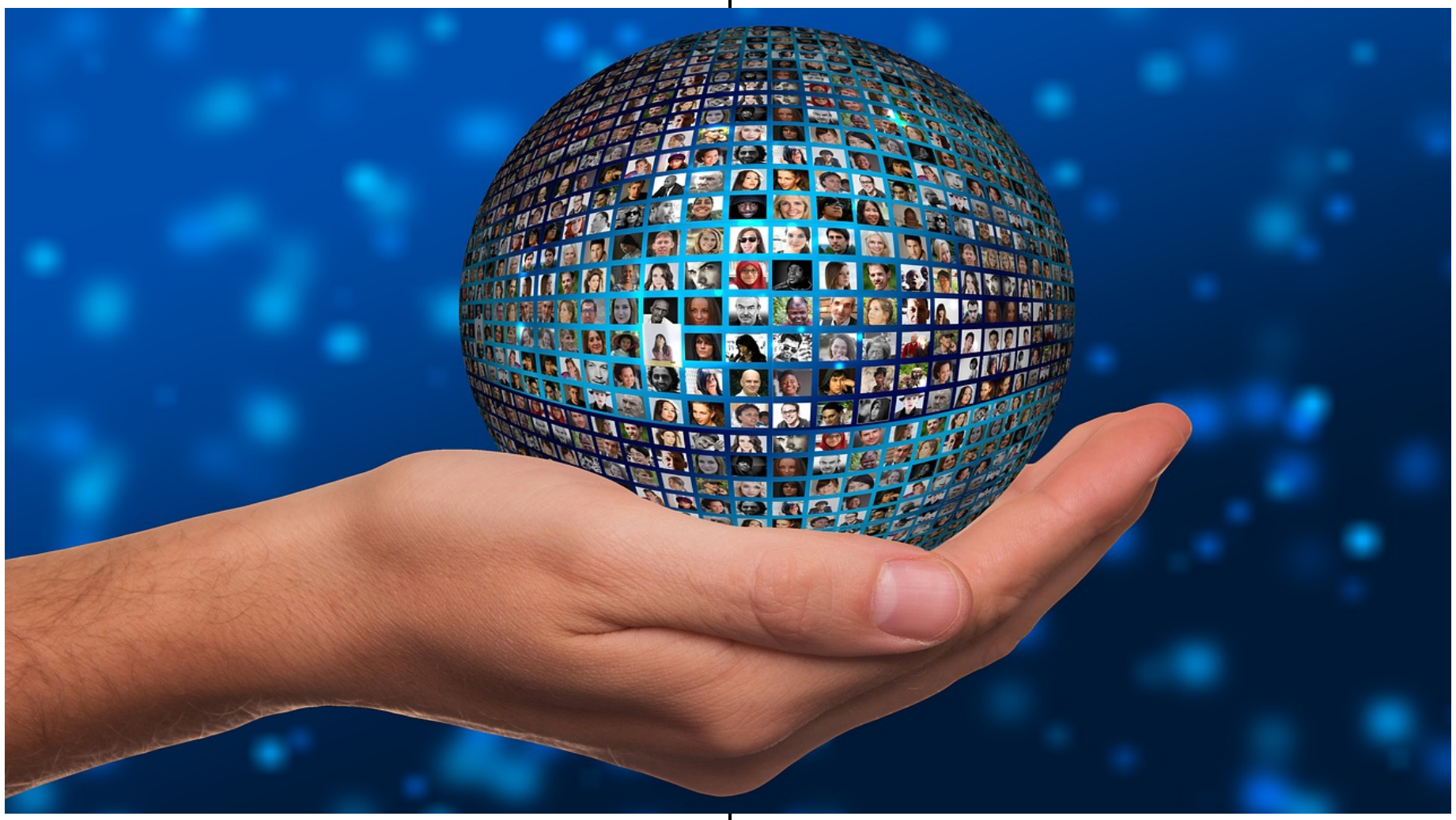

In this article we want to elicit the major lessons we have learned through Web Science and make some cautious predictions of what to expect next.

## The early years of the Web: A global information cyberspace

The early Web was planned as a global information space, a public agora for sharing documents, images and other resources. The value of the Web as a global cyberspace derived from its being a resilient, distributed system on which anyone could be both publisher and/or reader. Germane to its value was the topology of the Web, hyperlinks would connect the pages and linking would help to search for and find the most valuable resources.

During the early years, the interest in the Web as an object of scientific investigation was mostly technical. It led to better search engine technologies, to an improved infrastructure for sharing data, and for novel multimedia experiences. The early Web affected intellectual property rights and it caused significant, disruptive changes for many businesses. For instance, the music industry has been forever changed by the Web and still grapples with the changes that have resulted from sharing of audio files both legally and illegally. Despite this rapid change, however, in the early to mid 1990s, the Web was still only a footnote in the business reports of most companies.

## The modern web: A personalized and social information space

The changing of the millennium ushered in the second decade of the Web's use – with an increasing use of new tools that turned the information consumer into an information *prosumer*, someone both *pro*ducing and con*suming* content. In addition, as the Web grew, the increasing information being shared by users, both with and without their knowledge, paved the way for a *personalized* Web and the emergence of social networks. In the early global information cyberspace it was not considered important to track who visited a Web page or what they did there. However, with the growth of online commerce and the increasing spread of social networks, actions taken by users are massively tracked. Our browsers, and later cell phones, let companies that controlled web commerce know where we surfed to, what social media sites we use and who our friends are and details of our personal lives. With the increasing presence of smartphones, apps now record where we are, and report all of our clicks, likes, friendship networks, and whereabouts – information that is used to direct advertising, make business decisions, and otherwise use our online behaviors to know more about our social relationships, health, finances, and other aspects of our lives not directly reflected in our use of the Web.

[1] For example, in the US it is now reported that between 15-20% of newly married couples met their spouses on line (cf. http://www.statisticbrain.com/online-dating-statistics/).

The social Web goes far beyond linking documents. It started with linking of data (“what is your name?”, “where do you live?”, “who are your friends?”) to linking of actions (people who like this, tend to buy that), devices (e.g. smartphones) and people. It matters who creates such personal links, who owns such personal links (Google, Facebook, Tripadvisor, TripIt, etc.), and who is given access to such links (NSA, GCHQ, FSB, etc.). The power structure of the personalized and social Web is entirely different from the power structure of the early Web, as nowadays increasingly a small number of powerful companies control the data and often – though not always - collaborate with national governments, even if circumstances suggest collaboration to be illegal or unethical. This raises concerns about our democractic societies. The aforementioned article in 2006, already expressed the concern that “transparency and control over the complex social and legal relationships behind this information is vital”. For instance, the legal procedure against Facebook[2] from 2011 highlights that data which a user might reasonably be expected to be deleted may persist on the servers of a company, even against a user’s intention. Eduard Snowden revealed how mass surveillance intruded the privacy of basically every internet user and even included tracking information on governments between allied countries and, very recently, even their electoral processes.

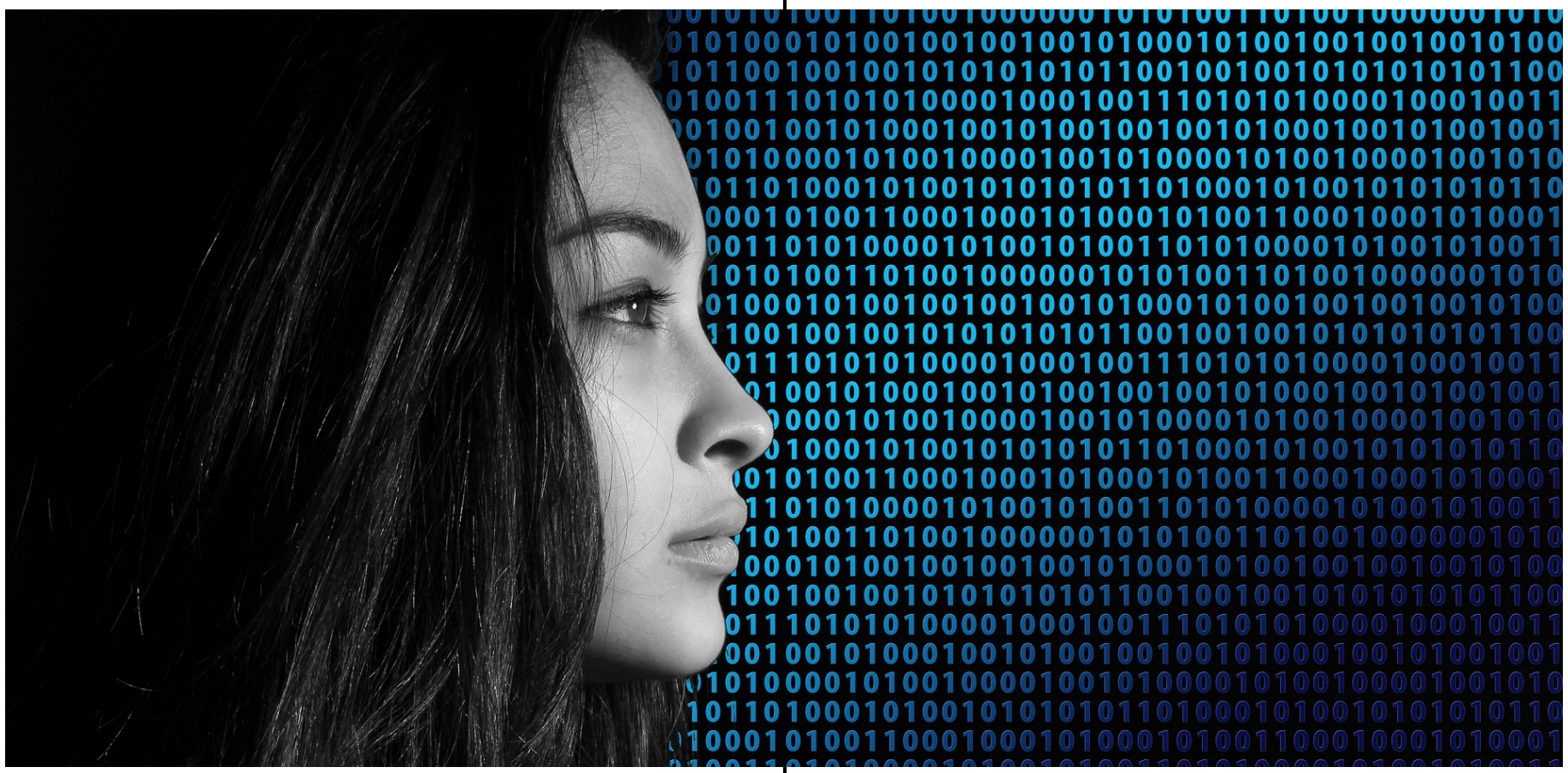

[2] Cf. http://europe-v-facebook.org/

Issues of personalization have also boosted the economic champions of the personalized Web who tailored physical and digital goods and services to the needs of consumers and who sold consumers’ attention to the attention seekers, such as advertisement networks. As personalization improves through the analysis of more and more data, the power structures of these powerful companies are reinforced. As a result, their roles are not only limited to economic control, as they increasingly begin to exert levels of control that were previously only attributed to governments. Power players like Facebook and Google decide within their “oligopoly” what may be published or what may be monetized. For instance, YouTube producers may monetize their content benefitting from thousands and sometimes millions of viewers – unless it includes political reporting or commentary for which YouTube, owned by Google, does not remunerate its producers at all.

## Web Science, the emerging discipline

To study this new global cyberspace, the field of Web Science has come into being to explore many different aspects of the Web, both its current use and the nature of its emergence [O’Hara et al, 2013]. Much of the interest in data science today arises because of the vast stores of information, both structured and unstructured, that have become available through the Web. The growth of available information is also leading to increasing use of data analytics in many fields, and the intersection of network-, data- and web-sciences is helping to bring new technologies to scientists and engineers working on large scale problems.

## The Web Observatory

One of the goals of Web Science is to be able to track and explore trends and usages of the information space that abounds. To this end, a number of of research groups in Web Science laboratories around the world have begun a project called the “Web Observatory” to collect and share data about the Web and its use [Tiropanis et al, 2013]. Increasingly, the move is from static analyses to tracking change in real-time, and to improve predictive models for understanding the impacts of information use across the network. Repository metadata and lightweight standards have been developed, and information is now being tracked from twelve countries on five continents. The continued development of a global Web observatory and data repository is enabling researchers to track and analyze past patterns of web usage and growth, and in the future forecasting models may help us better understand the impacts of emerging technologies.

## Trust and Social science

Web Science also investigates social problems that are propagated and exacerbated by the personalized and social Web. In the Web, trust is not only derived from actual experiences, but also from prejudice, leading to discrimination of people offering accomodation [Edelman & Luca, 2014], or to the effects of an echo chamber where one only receives partisan information based on previous choices [Coleoni et al, 2014], or feeding the frenzy of untruthful rumors [Friggeria et al, 2014]. Social media analysts look to understand, mathematically and socially, the trends being seen on the Web as reflected through information shared on social networking web sites and mobile applications. This is particularly relevant in current times as analysts pour over the social media data that arise from the recent Brexit referendum and the US presidential election.

## Social Machines

Researchers are also studying new phenomena that have arisen as a result of the co-constitution of the Web by people and machines including crowd-sourcing, collective intelligence and citizen science, and looking specifically at what can be learned from the collected data these technologies can generate. Web sites such as Wikipedia are powered by an interaction of many people, in many roles, and increasingly we see the growth of citizen science sites, like the "zooniverse" of scientific applications derived from the astronomical GalaxyZoo site [Lintott et al, 2008, Simpson et al, 2014]. These sites, harnessing the cognitive capacities of many millions of people, create powerful "social machines." Developing the principles for the successful design and governance of these sites, and lowering the barriers of entry for scientists and others in creating them, remains an active area of Web Science research [Hendler & Berners-Lee, 2010, Shadbolt et al, 2013].

## Big Data, Artificial Intelligence, and the World Wide Web

Over the past decade, new uses of the Web in both scientific and public discourse have grown, and shared data on the Web increasingly has become both a critical research resource and a challenge to manage technically and socially. The information available through mobile web platforms, such as geolocation information from smart phones, now powers a growing segment of new industries such as uber, lyft, airBnB and other parts of what has become known as the sharing economy. Governments in cities and countries around the world now release data on the Web in open formats [Janssen et al, 2013], open publishing is increasingly making journal and conference papers freely available to researchers, and online forums, such as PatientsLikeMe.com, provide new sources of information that researchers can tap. The emerging Internet of Things promises to yield even more data including much real-time data on the movement not just of people, but of energy and other resources needed for modern society.

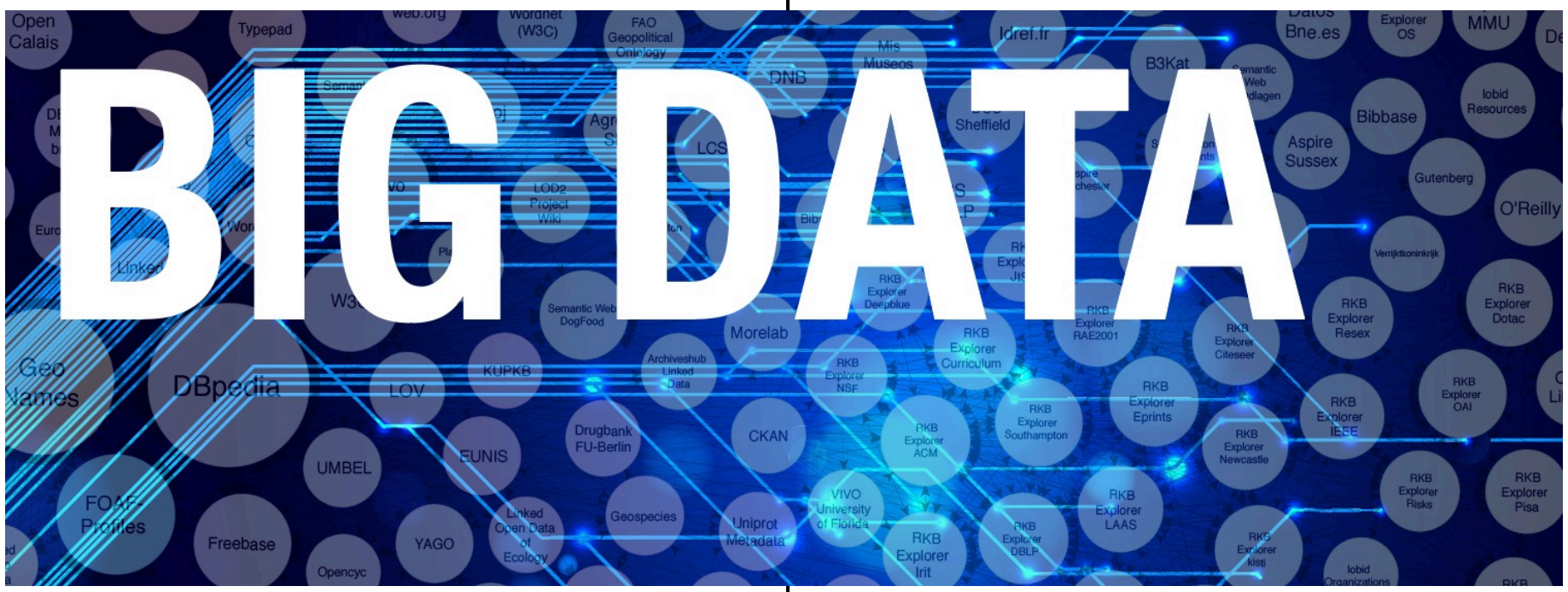

Through these developments, Artificial Intelligence (AI) and systems using AI are soaring. AI systems are used by physicians to guide diagnoses, by law firms to advise clients, by financial institutions to help decide who should receive loans, and by employers to guide whom to hire. The Web has been a prerequisite to these and further realms of automation, as it has enabled the collection of Big Data that is required to train these systems in the first place. The Web will also remain pivotal to keep these AI systems up to date with new data arriving from and being linked to droves of new digital devices on the emerging Internet of Things - from people and machines (cars, smart cities, etc.) from all over the world. However, the intelligence of these systems cannot be separated from the Web being a socio-technical system.

The provenance of Web data in automation reiterates all the issues encountered in the personalized and social Web – and more. Web data includes (i) data with technical and historic fallacies, (ii) non-representative samples of people, and (iii) subjective and biased decisions. Thus, AI systems that learn from Web data may find patterns that benefit decision making, at the same time they may not eliminate human biases from the decision-making process, but rather repeat and reinforce them. Thus, automation may lead to unintentional emergent properties of AI systems that "*deny historically disadvantaged and vulnerable groups full participation in society*" [Barocas & Selbst, 2016]. As Crawford & Calo [Crawford & Calo, 2016] remark, AI research has a blind spot that needs to be addressed by a social system approach to automated decision making, where the AI system undergoes a test of how the system is interwoven with social processes,ethical principles and legal regulations [Carmichael, 2016]. Indeed, this interdisciplinary approach has become a hallmark of Web Science research and is reflected in the growing number of social scientists, humanities scholars, political scientists and legal researchers at the various workshops, symposia, and conferences in the new discipline.

The Web is not limited to being a benefactor of Big Data and AI technologies. Rather, many of these technologies target the automation of processes in the Web. In terms of the economy, this kind of automation is more pervasive than the success of champions in the past Web. Under the heading "*The battle is for the customer interface*", Goodwin (2015) writes "*Uber, the world's largest taxi company, owns no vehicles. Facebook, the world's most popular media owner, creates no content. Alibaba, the most valuable retailer, has no inventory. And Airbnb, the world's largest accommodation provider, owns no real estate.*" While it may be true, that owning the primary interface maximizes returns, when it comes to customer retention that interface will need to come with as much intelligence as there could be – otherwise it might follow early and now forgotten success stories, such as Alta Vista or Lycos. The one who owns the most intelligent and automated interface may collect the most data with the highest representativeness and overall quality and may thus marginalize both competitors as well as workers who perform these – often non-trivial – jobs in many companies.

## Where next?

The evolution of the Web has resembled a land grab for the most promising homesteads in the Web, where whoever comes first has the best chances to define the rules of the space. Not only the Web itself, but even many services in the Web (social networks, search engines, electronic markets, etc.), constitute an infrastructure with network effects that favor the larger over the smaller Web platform. The Web thrives from competition, but it must not be throttled by monopolistic policies that often emerge in shared common infrastructure or in services where network effects dominate.

The Web and its services are becoming gatekeepers for the ways we communicate, live, play and work, and for our health and general well-being. Hence, code becomes law, but the law should not be imposed by the few without the control, or at least the knowledge, of the many. Regulatory principles, some of which are well-known for physical infrastructures such as roads and rails, need to be applied to the Web, e.g. granting access to core services in a non-discriminatory manner, which would not only remunerate entertainment, but also the political voices. It is critical that web scientists study, and communicate, not just the inherent promise, but also the risks and challenges the evolving Web will present [Hendler & Mulvehill, 2016].

The Web also distorts public voices, amplifying some, silencing others, eliciting the good, but also the bad and the ugly. Ways to dampen the hateful voices and encourage productive discussion clearly need to be found. Interestingly as polls are increasingly undermined as ways of forecasting the results of democratic elections, analysis of social media conversations is proving to be a more accurate methodology for such forecasts, despite the fact that social media users do not present a representative sample of the total electorate. Such results are still preliminary – much more research is needed – but maybe it is because the social networks act as echo chambers and give a larger majority, not heard in the controlled media a voice. Alternatively, it may be that the causality is the opposite – the echo chamber effect of information bubbles allows the propagation of distorted news and unfounded rumor. That is, we still need to study when social media represents a new republic with a greater voice and when it becomes the causal agent of manipulation of public opinion.

## Conclusion

All of these challenges – and more – need to be addressed in an ongoing socio-technical discussion to help us envision the future of the Web. New Web technology gives us unprecedented means to capture, analyze and benefit from Big Data and Big Knowledge, such as personalized medicine or smarter travel. The Web mindset gives us headway for unprecedented collaboration – be it Open Science, Open Education or Digital Democracy. The emergent Internet of Things and the application of AI and blockchain technologies promises much in terms of smarter everything but we can also see a nightmare world of control by a network of machines and devices that we have little control over. All these promises need to be framed within understanding and public discussions of the implied social systems. Ten years ago we argued that privacy, security and trust on the Web would be amongst the most important things for Web Scientists to research. Our work in this respect has only just begun.

The Web has undergone a development from a public space of documents towards automated, personalized transactions and highly connected social networks. It started with linked documents and linked data and is now linking services, things (sensors and actors) and people. We are at the point, where the Web and what is linked to the Web is the prerequisite for automation, the means of automation and the objective of automation.

The future of the Web is by no means certain. There are many pressures on it both social and technical (cyber crime, cyber security, commercial, geo-political, alternative internet architectures) which could cause it to fragment, and internet governance is an on-going matter of urgent public debate. The future of the Web is inextricably tied to the outcome of these discussions. Over the next ten years will the Web become a dark, fragmented and anti-social place, or a platform fo collaboration, prosperity and development?

The Web is an ongoing experiment, because we do not know and cannot extrapolate how it's development will change the fabric of our government, our society, our work and our lives. What we do know is that we need to observe, understand and discuss how the Web develops, because it is a development that can and should serve for the best of humanity, but it might fall prey to the worst. This work must be done from a socio-technical perspective, which makes Web Science even more important now than it was when we launched it ten years ago.